\documentclass{tlp_corr}

\pubyear{2011}
\submitted {7th October 2009}
\revised {15th March 2010}
\accepted{13th Feb 2011}

\usepackage{times}

\usepackage{graphicx}
\usepackage{listings}
\usepackage{color}
\usepackage{amssymb}
\usepackage{amsmath}
\usepackage{stmaryrd}

\usepackage{tikz}
\usetikzlibrary{positioning,shapes,arrows,fit,shadows,calc,matrix}

\usepackage[latin1]{inputenc}

\definecolor{lightgrey}{rgb}{0.95,0.95,0.95}
\newcommand{\ciaopp}{CiaoPP}
\newcommand{\ciao}{Ciao}
\newcommand{\yap}{YAP}
\newcommand{\bprolog}{B-Prolog}
\newcommand{\xsb}{XSB}
\newcommand{\sicstus}{SICStus Prolog}

\newcommand{\swiprolog}{SWI-Prolog}
\newcommand{\gnuprolog}{GNU Prolog}
\newcommand{\binprolog}{BinProlog}
\newcommand{\quintus}{Quintus Prolog}

\newcommand{\codehlt}[1]{\colorbox{lightgrey}{\small\texttt{#1}}}
\newcommand{\codehltw}[1]{\colorbox{lightgrey}{\parbox{\textwidth}{\small\texttt{#1}}}}

\newcommand{\reducespace}{}
\newcommand{\negskip}{\vspace*{-1.7mm}}
\newcommand{\figskip}{\vspace*{-3mm}}

\newcommand{\parbegin}{\vspace*{-3.5mm}}
\newcommand{\underfig}{\vspace*{-1mm}}

\newcommand{\nt}[1]{\mbox{\it #1}}

\newcommand{\kbd}[1]{\mbox{\tt #1}}

\newtheorem{example}{Example}

\title[
An Overview of \ciao\ and its Design Philosophy
]
{
An Overview of \ciao\ and its Design Philosophy
}
\author[M.V. Hermenegildo et al.]{
   M. V. HERMENEGILDO$^{1,2}$  
   ~~  F. BUENO$^1$  
   ~~ M. CARRO$^1$  \and
   ~~ P. L\'{O}PEZ-GARC\'{I}A$^{2,4}$  
   ~~ E. MERA$^3$  
   ~~ J. F. MORALES$^2$ 
\vspace*{2mm}
   ~~ G. PUEBLA$^1$\\ 
   $^1$Universidad Politécnica de Madrid (UPM) \\
\email{bueno@fi.upm.es, mcarro@fi.upm.es, german@fi.upm.es} \\
   $^2$Madrid Institute of Advanced Studies in Software Development
   Technology (IMDEA Software)\\
\email{manuel.hermenegildo@imdea.org, pedro.lopez@imdea.org, josef.morales@imdea.org}\\
   $^3$Universidad Complutense de Madrid (UCM)\\
\email{edison@fdi.ucm.es}\\
   $^4$Spanish Research Council (CSIC) \\
\vspace*{-5mm}
}

\begin{document}

\maketitle

\begin{abstract}
  We provide an overall description of the \ciao\ multiparadigm
  programming system emphasizing some of the novel aspects and
  motivations behind its design and implementation.  An important
  aspect of \ciao\ is that, in addition to supporting logic
  programming (and, in particular, Prolog), it provides the programmer
  with a large number of useful features from different programming
  paradigms and styles, and that the use of each of these features
  (including those of Prolog) can be turned on and off at will for
  each program module. Thus, a given module may be using, e.g., higher
  order functions and constraints, while another module may be using
  assignment, predicates, Prolog meta-programming, and concurrency.
  Furthermore, the language is designed to be extensible in a simple
  and modular way.
  Another important aspect of \ciao\ is its programming environment,
  which provides a powerful preprocessor (with an associated assertion
  language) capable of statically finding non-trivial bugs, verifying
  that programs comply with specifications, and performing many types
  of optimizations (including automatic parallelization).
  Such optimizations produce code that is highly competitive with
  other dynamic languages or, with the (experimental) optimizing
  compiler, even that of static languages, all while retaining the
  flexibility and interactive development of a dynamic language.
  This compilation architecture supports modularity and
  separate compilation throughout. 
  The environment also includes a powerful auto-documenter and a unit
  testing framework, both closely integrated with the assertion system.
  The paper provides an informal overview of the language and program
  development environment.  It aims at illustrating the design
  philosophy rather than at being exhaustive, which would be
  impossible in a single journal paper, pointing instead to previous
  \ciao\ literature.
\end{abstract}


\section{Origins and Initial Motivations}
\label{sec:orig-init-motiv}

\begin{figure}[t]
  \centering
  \input{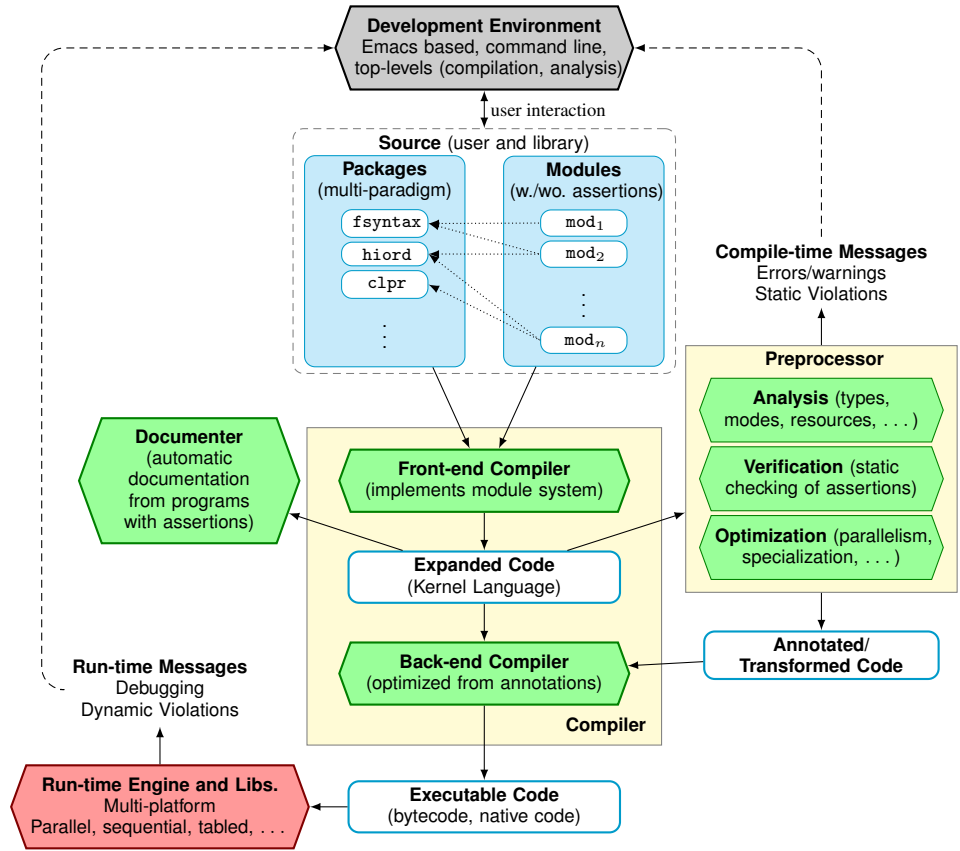}
  \figskip
  \caption{A high-level view of the \ciao\ system.}
  \label{fig:ciao-overall}
\underfig
\underfig
\end{figure}

\ciao~\cite{ciao-ppcp-short,ciao-novascience-short,ciao-reference-manual-1.13-short,ciao-philosophy-note-tr-short}
is a modern, multiparadigm programming language with an advanced
programming environment.  The ultimate motivation behind the system is
to develop a combination of programming language and development tools
that together help programmers produce in less time and with less
effort code that has fewer or no bugs.  \ciao~aims at combining the
flexibility of dynamic/scripting languages with the guarantees and
performance of static languages. It is designed to run very
efficiently on platforms ranging from small embedded processors to
powerful multicore architectures.
Figure~\ref{fig:ciao-overall} shows an overview of the \ciao\ system
architecture and the relationships among its components, which will be
explained throughout the paper.

\ciao\ has its main roots in the \&-Prolog language and system
\cite{ngc-and-prolog}.  \&-Prolog's design was aimed at
achieving higher performance than state of the art sequential logic
programming systems by exploiting parallelism, in particular,
and-parallelism~\cite{sinsi-jlp-short}. This required the development
of a specialized abstract machine, derived from early versions of
SICStus Prolog~\cite{sicstus-manual-4.1.1}, capable of running a large
number of (possibly non-deterministic) goals in
parallel~\cite{Hampaper-short,ngc-and-prolog}. The source
language was also extended in order to allow expressing parallelism
and concurrency in programs, and later to support constraint
programming, including the concurrent and parallel execution of such
programs~\cite{clppar-plilp96-short}.

Parallelization was done either by hand or by means of the \&-Prolog
compiler, which was capable of automatically annotating programs for
parallel
execution~\cite{iclp90-annotation-short,annotators-jlp-short}.  This
required developing advanced program analysis technology based on
abstract interpretation~\cite{Cousot77-short}, which led to the
development of the PLAI
analyzer~\cite{pracabsin-short,pracai-jlp-short,ai-jlp-short}, based
on Bruynooghe's approach~\cite{bruy91} but using a highly-efficient
fixpoint including memo tables, convergence acceleration, dependency
tracking, etc.  This analyzer inferred program properties such as
independence among program
variables~\cite{ai-jlp-short,freeness-iclp91-short}, absence of side
effects, non-failure ~\cite{nfplai-flops04-short},
determinacy~\cite{determinacy-ngc09-short}, data structure shape and
instantiation state (``moded
types'')~\cite{Saglam-Gallagher-95-short,eterms-sas02-short}, or upper
and lower bounds on the sizes of data structures and the cost of
procedures~\cite{granularity-short,caslog-short,low-bounds-ilps97-short}.
This was instrumental for performing automatic granularity
control~\cite{granularity-short,granularity-jsc-short}.  In addition
to automatic parallelization the \&-Prolog compiler performed other
optimizations such as multiple (abstract)
specialization~\cite{spec-pepm-short}.  Additional work was also
performed to extend the system to support other computation rules,
such as the Andorra principle~\cite{Warren93-short,pl2aI-lopstr-short}
and other sublanguages and control rules.

In the process of gradually extending the
capabilities of the \&-Prolog system in the late 80's/early 90's
two things became clear.
Firstly, the wealth of information inferred by the analyzers would
also be very useful as an aid in the program development process.
This led to the idea of the \ciao\ assertion
language and preprocessor,
two fundamental components of the \ciao\ system (even if neither of
them are strictly required for developing or compiling programs).
The \ciao\ assertion language~\cite{assert-lang-disciplbook-short}
provides a homogeneous framework which allows, among other things,
static and dynamic verification to work cooperatively in a unified
way.
The \ciao~Preprocessor
(\ciaopp~\cite{prog-glob-an-short,preproc-disciplbook-short,ciaopp-sas03-journal-scp-short})
is a powerful tool capable of statically finding non-trivial bugs,
verifying that the program complies with specifications (written in
the assertion language), and performing many types of program
optimizations.

A second realization was that many desirable language extensions
could be  supported efficiently within 
the same system if the underlying machinery implemented a relatively
limited set of basic constructs (a \emph{kernel
  language})~\cite{ciao-ppcp-short,ciao-novascience-short} coupled
with an easily programmable and modular way of defining new syntax and
giving semantics to it in terms of that kernel language.
This idea is not exclusive to \ciao, but in \ciao\ the facilities that
enable building up from a simple kernel are explicitly available from
the system programmer level to the application programmer level.  The
need to be able to define extensions based on some basic blocks led to
the development of a novel module
system~\cite{ciao-modules-cl2000-short} which allows writing language
extensions (\emph{packages}) by grouping together syntactic
definitions, compilation options, and plugins to the compiler.
The mechanisms provided for adding new syntax to
the language and giving semantics to such syntax can
be activated or deactivated on a per-compilation unit basis without
interfering with other units. 
As a result all \ciao\ operators, ``builtins,'' and most other syntactic
and semantic language constructs are
user-modifiable and live in \emph{libraries}.%
\footnote{In fact, some \ciao\ packages are portable with little
  modification to other logic and constraint logic programming
  systems. Others require support from the kernel language
  (e.g. concurrency), to provide the desired semantics or
  efficiency. In any case, packages offer a \emph{modularized} view of
  language extensions to the user.}  The \ciao\ module system also
addresses the needs for modularity deriving from global analysis.
We will start precisely with the introduction of the user view of
packages.

\section{Supporting Multiple Paradigms and Useful Features}
\label{sec:supp-mult-parad}

Packages allow \ciao\ to support multiple programming paradigms and
styles in a single program.
The different source-level sub-languages are supported by a
compilation process stated by the corresponding package, typically via
a set of rules defining \emph{source-to-source} transformations into
the kernel language.
This kernel is essentially pure Prolog plus a number of basic,
instrumental additional functionalities (such as the cut, non-logical
predicates such as \texttt{var/1} or \texttt{assert/1}, threads,
attributed variables, etc.), all of which are in principle not visible
to the user but can be used if needed at the kernel level to support
higher-level functionality.  However, the actual nature of the kernel
language is actually less important than the extensibility mechanisms
which allow these extensions to be, from the point of view of the
compiler, analyzers, autodocumenter, and language users, on a par with
the native builtins.
We will now show some examples of how the extensibility provided by
the module system allows \ciao\ to incorporate the fundamental
constructs from a number of programming paradigms.

\begin{figure}[t]
\begin{lstlisting}
:- module(_, _, [functional, lazy]). (*@\label{functions}@*)

nrev([])    := []. (*@\label{nrev}@*)
nrev([H|T]) := ~conc(nrev(T), [H]).

conc([],    L) := L. (*@\label{conc}@*)
conc([H|T], K) := [H | conc(T, K)]. 

fact(N) := N=0 ? 1 (*@\label{fact}@*)
	|  N>0 ? N * fact(--N).

:- lazy fun_eval nums_from/1. (*@\label{numsfrom}@*)
nums_from(X) := [X | nums_from(X+1)].

:- use_module(library('lazy/lazy_lib'), [take/3]). (*@\label{nums}@*)
nums(N) := ~take(N, nums_from(0)).
\end{lstlisting}
\figskip
\caption{Some examples in \ciao\ functional notation.}
\label{fig:fnrev}
\underfig
\end{figure}

We will use the examples in Fig.~\ref{fig:fnrev} to illustrate
general concepts regarding the module system and its extensibility.
In \ciao\ the first and second arguments of a \texttt{module}
declaration (line \ref{functions}) hold the module name and list of
exports in the standard way.  ``\texttt{\_}'' in the first argument
means that the name of the module is the name of the file, without
suffix, and in the second one that all definitions are exported.
The third argument states  a list of packages to
be loaded (\texttt{functional} and \texttt{lazy} in this case, which
provide functional notation and lazy evaluation).  
Packages are \ciao~files which contain syntax and compilation rules and
which are loaded by the compiler as plugins and unloaded when
compilation finishes.  Packages only modify the syntax and semantics
of the module from where they are loaded, and therefore other modules
can use packages introducing incompatible syntax / semantics without
clashing.  Packages can also be loaded using \texttt{use\_package}
declarations throughout the module.

\parbegin
\paragraph{\textbf{Functional Programming:}} 
functional notation~\cite{functional-lazy-notation-flops2006-short} is
provided by a set of packages which, besides a convenient syntax to
define predicates using a function-like layout, gives support for
semantic extensions which include higher-order facilities (e.g.,
predicate abstractions and applications thereof) and, if so required,
lazy evaluation.
Semantically, the extension is related to logic-functional languages
like Curry~\cite{Hanus03curry} but relies on flattening and
resolution, using \texttt{freeze/2} for lazy evaluation, instead of
narrowing.
For illustration, Fig.~\ref{fig:fnrev} lists a number of examples
using the \ciao\ functional notation.  Thanks to the packages
loaded by the \texttt{module} declaration, \texttt{nrev} and
\texttt{conc} can be written in functional style by using multiple
\texttt{:=/2} definitions.  The \verb+~+ prefix operator in the second
rule for \texttt{nrev} states that its argument (\texttt{conc}) is an
interpreted function (a call to a predicate), as opposed to a data
structure to unify with and return as a result of function invocation.
This \emph{eval} mark can be omitted when the predicate is 
marked for functional syntax.  The recursive call to
\texttt{nrev} does not need such a clarification because it is called
within its own definition.
The list constructor in \texttt{conc} is not marked for evaluation,
and therefore it stands for a data structure instead of a predicate
call.

\texttt{fact} is written using a disjunction (marked by
``\texttt{|}'') of guards (delimited by ``\texttt{?}'')  which
together commit the system to the first matching choice.  Arithmetic
operators are assumed to be evaluable by default, but this can be
turned off with a special declaration.  \texttt{nums\_from} is
declared lazy, which makes it possible to write a recursion which is
executed only up to the extent it is necessary.  In this case, it is
called by \texttt{take} (imported from a library of lazy
functions/predicates) which in turns allows \texttt{nums} to (lazily)
return a list of \texttt{N} numbers starting at \texttt{0}.

The following queries produce the expected answer:
\negskip
\begin{small}
\begin{verbatim}
?- use_package(functional).
?- X = ~nrev([1,2,3]).
   X = [3,2,1]
?- [3,2,1] = ~nrev(X).
   X = [1,2,3]
\end{verbatim}
\end{small}
\negskip
Loading the \texttt{functional} package in the top level allows using
functional notation in it ---the top level behaves in this sense essentially
in the same way as a module.  Since in general, functional
\emph{notation} is just syntax and thus no directionality is implied,
the second query to \texttt{nrev/2} just instantiates its argument.

However, as mentioned before, other constructs such as conditionals do
commit the system to the first matching case.  The assertion language
includes \texttt{func} assertions aimed at enforcing strictly
``functional'' behavior (e.g., being single moded, in the sense that a
fixed set of inputs must always be ground and for them a single output
is produced, etc.), and generating \emph{assertions} (see later) which
ensure that the code is used in a functional way.

\begin{figure}[t]
\begin{lstlisting}[name=fmore]
:- module(someprops, _, [functional, hiord]).

color := red | blue | green.

list := [] | [_ | list].

list_of(T) := [] | [~T | list_of(T)].

sorted := [] | [_].
sorted([X,Y|Z]) :- X @< Y, sorted([Y|Z]).
\end{lstlisting}

\begin{lstlisting}[name=fmore,firstnumber=1]
:- module(someprops, _, []).

color(red).   color(blue).   color(green).

list([]).
list([_|T]) :- list(T).

:- use_module(engine(hiord_rt)).

list_of(_, []).
list_of(T, [X|Xs]) :- call(T, X), list_of(T, Xs).

sorted([]).    sorted([_]).
sorted([X,Y|Z]) :- X @< Y, sorted([Y|Z]).
\end{lstlisting}
\figskip
\caption{Examples in \ciao\ functional notation and state of translation
  after applying the \texttt{functional} and \texttt{hiord} packages.}
\label{fig:fmore}
\underfig
\end{figure}

Fig.~\ref{fig:fmore}
lists more examples using \texttt{functional} and other packages, and
the result after applying just the transformations brought in by the
\texttt{functional} package. Note the use of higher order in
\texttt{list\_of}: a predicate is called using a syntax which has a
variable in the place of a predicate name.  This is possible thanks to
the \texttt{hiord} package (more on it later) which adds the necessary
syntax and a 
compile-time translation into \texttt{call/N}.  

\parbegin
\paragraph{\textbf{Classic and ISO-Prolog:}} 
\ciao\ provides, through convenient defaults, an excellent Prolog
system with support for ISO-Prolog.  Other classical ``builtins''
expected by users, and which are provided by modern Prolog systems
(\yap, \swiprolog, \quintus, \sicstus, \xsb, \gnuprolog, \bprolog,
\binprolog, etc.), are also conveniently available.  In line with its
design philosophy, in \ciao\ all of these features are optional and
brought in from libraries rather than being part of the language.
This is done in such a way that classical Prolog code runs without
modifications: the Prolog libraries are automatically loaded when
module declarations have only the first two arguments, which is the
type of module declaration used by most Prolog systems (see
Fig.~\ref{fig:prolog}, left).  This is equivalent to loading only the
``\texttt{classic}'' package (Fig.~\ref{fig:prolog}, right).

\begin{figure}[t]
  \begin{minipage}{0.46\linewidth}
\begin{lstlisting}[name=prolog1,numbers=none]
:- module(h,[main/1]).

main :- write("Hello world!").
\end{lstlisting}
  \end{minipage}
\hfill
  \begin{minipage}{0.50\linewidth}
\begin{lstlisting}[name=prolog2,numbers=none]
:- module(h,[main/1],[classic]).

main :- write("Hello world!").
\end{lstlisting}
  \end{minipage}
\figskip
\caption{Two equivalent Prolog modules.}
\label{fig:prolog}
\underfig
\end{figure}
    
The set of ISO builtins and other ISO compliance-related features
(e.g., the exceptions they throw) are triggered by loading the
\texttt{iso} package (included in \texttt{classic}).  Facilities for
testing ISO compliance (Section~\ref{sec:unit-testing}) are also
available.

The \texttt{classic} Prolog package is also loaded by default in user
files (i.e., those without a \texttt{module} declaration) that do not
load any packages explicitly via a \texttt{use\_package} declaration.
Also, the system top level comes up by default in Prolog mode.  This
can be tailored by creating a \verb+~/.ciaorc+ initialization file
which, among other purposes, can be used to state packages to be
loaded into the top level. As a result of these defaults, \ciao\ users
who come to the system looking for a Prolog implementation do get what
they expect.  If they do not poke further into the menus and manuals,
they may never realize that \ciao\ is in fact quite a different beast
under the hood.

\parbegin
\paragraph{\textbf{Other Logic Programming Flavors:}} 
alternatively to the above, by not loading the classic Prolog
package(s) the user can restrict a given module to use only pure logic
programming, without any of Prolog's impure features.\footnote{The
  current implementation --as of version 1.13-- does still leave a few
  builtins visible, some of them useful for debugging. To avoid the
  loading of any impure builtins in 1.13 the \texttt{pure}
  pseudo-package should be used.}  That means that if a call to
\texttt{assert} were to appear within the module, it would be signaled
by the compiler as a call to an undefined predicate.  Features such
as, for example, declarative I/O, can be added to such pure modules by
loading additional libraries. This also allows adding individual
features of Prolog to the pure kernel on a needed basis.

\emph{Higher-order logic programming} with \emph{predicate
  abstractions} (similar to \emph{closures}) is supported through
the~\texttt{hiord} package. This is also illustrated in
Fig.~\ref{fig:fmore}, where the \texttt{list\_of/2} predicate receives
a unary predicate which is applied to all the arguments of a list.  As
a further example of the capabilities of the~\texttt{hiord} package,
consider the queries:
\begin{verbatim}
    ?- use_package(hiord), use_module(library(hiordlib)).
    ?- P = ( _(X,Y) :- Y = f(X) ),    map([1, 3, 2], P, R).
\end{verbatim}
\noindent
where, after loading the higher-order package \texttt{hiord} and
instantiating \texttt{P} to the anonymous predicate \texttt{\_(X,Y) :-
  Y = f(X)}, the call \texttt{map([1, 3, 2], P, R)} applies \texttt{P}
to each element of the list \texttt{[1, 3, 2]} producing
\verb'R = [f(1), f(3), f(2)]'.
The (reversed) query works as expected, too:
\begin{verbatim}
    ?- P = ( _(X,Y) :- Y = f(X) ),    map(M, P, [f(1), f(3), f(2)]).
       M = [1, 3, 2]
\end{verbatim}

If there is a free variable, say \texttt{V}, in the predicate
abstraction and a variable with the same name \texttt{V} in the clause
within which the anonymous predicate is defined, the variable in the
predicate abstraction is bound to the value of the variable in the
clause.  Otherwise it is a free variable, in the logical sense (as any 
other existential variable in a clause).  This is independent from the
environment where the predicate abstraction is applied, and therefore
closures have syntactic scoping. 

\parbegin
\paragraph{\textbf{Additional Computation Rules:}} 
in addition to the usual depth-first, left-to-right execution of
Prolog, other computation rules such as breadth-first, iterative
deepening, tabling (see later), and the Andorra model are available,
again by loading suitable packages.  This has proved particularly
useful when teaching, since it allows postponing the introduction of
the (often useful in practice) quirks of Prolog (see the slides of a
course starting with pure logic programming and breadth-first search
in \texttt{\url{http://www.cliplab.org/logalg}}).

\begin{figure}[t]
\begin{lstlisting}
:- module(_,_,[fsyntax,clpqf]).

fact(.=. 0) := .=. 1. (*@\label{factbase}@*)
fact(N)     := .=. N*fact(.=. N-1) :- N .>. 0. 

sorted := [] | [_].
sorted([X,Y|Z]) :- X .<. Y, sorted([Y|Z]).
\end{lstlisting}
\figskip
\caption{\ciao\ constraints (combined with functional notation).}
\label{fig:fcons}
\underfig
\end{figure}

\parbegin
\paragraph{\textbf{Constraint Programming:}} 
several constraint solvers and classes of constraints using these
solvers are supported including CLP($\mathcal{Q}$), CLP($\mathcal{R}$)
(a derivative of~\cite{holzbaur-clone}), and a basic but usable
CLP($\mathcal{FD}$) solver.
\footnote{CLP($\mathcal{X}$) stands for a Constraint Logic Programing
  System parametrized by the constraint domain $\cal{X}$.}
The constraint languages and solvers, which are built on more basic
blocks such as attributed variables~\cite{holzbaur-plilp92-short}
and/or the higher-level Constraint Handling Rules
(CHR)~\cite{fruehwirth09:CHR_book}, also available in \ciao, are
extensible at the user level.

Fig.~\ref{fig:fcons} provides two examples using \ciao\
CLP($\mathcal{Q}$) constraints, combined with functional notation. For
example, line~\ref{factbase} can be read as: if the input argument of
\texttt{fact} is constrained to 0 then the ``output'' argument is
constrained to 1.  In the next line, if the argument of \texttt{fact}
is constrained to be greater than 0 then the ``output'' is constrained
to be equal to \texttt{N*fact( .=. N-1 )}.  The two definitions
(\texttt{fact} and \texttt{sorted}) can be called with their arguments
in any state of instantiation. For example, the query
\negskip
\begin{verbatim}
?- sorted(X).
\end{verbatim}
\negskip
\noindent
returns (blanks in the answers have been edited to save space):
\negskip
\begin{verbatim}
X = [] ? ;
X = [_] ? ;
X = [_A, _B], _A .<. _B ? ;
X = [_A, _B, _C], _B .<. _C, _A .<. _B ?
\end{verbatim}
\negskip
\noindent
etc.\ 
As many other CLP systems \ciao~is not, at the moment, a highly
specialized constraint system, and it does not intend to compete with
very high performance systems like, e.g.,
Gecode~\cite{SchulteStuckey:TOPLAS:2008} or
Comet~\cite{hentenryck05:comet}.  The purpose of the constraint
solving support present in \ciao~is to offer some reasonable
functionality for medium-sized problems and to be able to explore new
possibilities in the combination of paradigms.

\parbegin
\paragraph{\textbf{Object-Oriented Programming:} }
object oriented-style programming has been classically provided in
\ciao~through the O'\ciao~\texttt{class} and \texttt{object}
packages~\cite{pineda02:ociao-short}. These packages provide
capabilities for class definition, object instantiation, encapsulation
and replication of state, inheritance, interfaces, etc. These features
are designed to be natural extensions of the underlying module
system. There is current work performed within the
``\texttt{optimcomp}'' branch (see later) revisiting these issues in
the context of abstract mechanisms for passing, maintaining, and
updating different notions of state. These extensions have also
introduced imperative control structures and nested syntactic scopes.

\parbegin
\paragraph{\textbf{Concurrency, Parallelism, and Distributed Execution:}}
other packages bring in different capabilities for expressing
concurrency (including a concurrent, shared version of the internal
fact database which can be used for
synchronization~\cite{shared-database}), distribution, and parallel
execution~\cite{ciao-dis-impl-parimp-short,hlfullandpar-iclp2008}.  A
notion of ``active objects'' also allows compiling objects so that
they are ultimately mapped to a standalone process, which can then be
transparently accessed by the rest of an application. This provides
simple ways to implement servers and services in general.

\ \\ [-1mm]
\noindent
In addition to the programming paradigm-specific characteristics
above, many additional features are available through
libraries (that can also be activated or deactivated on a
\emph{per-module / class} basis), including:

\parbegin
\paragraph{\textbf{Structures with named arguments}}(feature terms), a
trimmed-down version of $\psi$-terms~\cite{HASSAN93} which translates
structure unifications to Prolog unifications, adding 
no overhead to the execution when argument names can be statically
resolved, and a small overhead when they are resolved at run time.

\parbegin
\paragraph{\textbf{Partial support for advanced higher-order}}logic
programming features, like higher-order unification, based on the
algorithms used in $\lambda$Prolog~\cite{Wolfram92} (experimental).

\parbegin
\paragraph{\textbf{Persistence},}which allows \ciao\ to transparently save
and restore the state of selected facts of the dynamic database of a
program on exit and startup.  This is the basis of a high-level
interface with databases~\cite{persdb-padl-2004-short}.

\parbegin
\paragraph{\textbf{Tabled evaluation}}\cite{chen96:tabled_evaluation}, pioneered by XSB (experimental).

\parbegin
\paragraph{\textbf{Answer Set Programming (ASP)}}\cite{asp-prolog},
which makes it possible to execute logic programs under the
\emph{stable model semantics} (experimental).

\parbegin
\paragraph{\textbf{WWW programming},} which establishes a direct
mapping of HTML / XML and other formats to Herbrand terms, allowing
the manipulation
of WWW-related data easily through unification, 
writing CGIs, etc.~\cite{pillow-tplp-short}.

\section{ \ciao\ Assertions}
\label{sec:assertions}

An important feature of \ciao\ is the availability of a rich,
multi-purpose assertion language. We now introduce (a subset of) this
assertion language.  Note that a great deal of the capabilities of
\ciao\ for supporting and processing assertions draws on its
extensibility features which are used to define and give semantics to
the assertion language without having to change the low-level
compiler.

\parbegin
\paragraph{\textbf{\ciao\ Assertion Language Syntax and Meaning:}}

Assertions are linguistic constructs which allow expressing properties
of programs.  Syntactically they appear as an extended set of
declarations, and semantically they allow talking about preconditions,
(conditional-) postconditions, whole executions, program points, etc.
For clarity of exposition, we will focus on the most commonly-used
subset of the \ciao~assertion language: \texttt{pred} assertions and
program point assertions.  A detailed description of the full language
can be found
in~\cite{assert-lang-disciplbook-short,ciao-reference-manual-1.13-short}.

\begin{figure}[t]
\begin{lstlisting}[name=fmoreprops]
:- module(someprops, _, [functional, hiord, assertions]).
:- prop color/1.    color := red | blue | green.
:- prop list/1.     list := [] | [_ | list].
:- prop list_of/2.  list_of(T) := [] | [~T | list_of(T)].
:- prop sorted/1.   sorted := [] | [_].
                    sorted([X,Y|Z]) :- X @< Y, sorted([Y|Z]).
\end{lstlisting}
\figskip
\caption{Examples of state \texttt{prop}erty definitions.}
\label{fig:fmoreprops}
\underfig
\end{figure}

The first subset, \texttt{pred} \textbf{assertions}, is used to
describe a particular predicate. They can be used to state
preconditions and postconditions on the (values of) variables in the
computation of predicates, as well as global properties of such
computations (such as, e.g., the number of execution steps,
determinacy, or the usage of some other resource).
Fig.~\ref{fig:nrevf} includes a number of \texttt{pred} assertions
whose syntax is made available through the \texttt{assertions}
package.  For example, the assertion (line~\ref{listuse}):
\ \hfill \codehlt{:- pred nrev(A,B) : list(A) => list(B).}\\
expresses that calls to predicate \texttt{nrev/2} with the first
argument bound to a list are admissible, and that if such calls
succeed then the second argument should also be bound to a list.
\texttt{list/1} is an example of a {\em state property} --a
\texttt{prop}, for short: a predicate which expresses properties of
the (values of) variables.  Other examples are defined in
Fig.~\ref{fig:fmoreprops} (\texttt{sorted/1}, \texttt{color/1},
\texttt{list\_of/2}), or arithmetic predicates such as \texttt{>/2},
etc.  Note that \texttt{A} in \texttt{list(A)} above refers to the
first argument of \texttt{nrev/2}.  We could have used the parametric
type \texttt{list\_of/2} (also defined in Fig.~\ref{fig:fmoreprops}),
whose first argument is a type parameter, and written
\texttt{list\_of(term,A)} instead of \texttt{list(A)}, where the type
\texttt{term/1} denotes any term.
As an additional example using the parametric type
\texttt{list\_of/2}, the assertion in line~\ref{colorlistuse} of
Fig.~\ref{fig:nrevf} expresses that for any call to predicate
\texttt{nrev/2} with the first argument bound to a list of
\texttt{color}s, if the call succeeds, then the second argument is
also bound to a list of \texttt{color}s.

State properties defined by the user and exported/imported as usual.
In Fig.~\ref{fig:nrevf} some properties (\texttt{list/1},
\texttt{list\_of/2}, \texttt{color/1}) are imported from the user
module \texttt{someprops} (Fig.~\ref{fig:fmoreprops}) and others
(e.g., \texttt{size\_ub/2}) from the system's \texttt{nativeprops}.
In any case \texttt{prop}s need to be marked explicitly as such (see
Fig,~\ref{fig:fmoreprops}) and this flags that they need to meet some
restrictions~\cite{assert-lang-disciplbook-short,ciao-reference-manual-1.13-short}.
E.g., their execution should terminate for any possible call since, as
discussed later, \texttt{prop}s will not only be checked at compile
time, but may also be involved in run-time checks.
Types are just a particular case (further restriction) of state
properties. Different type systems, such as regular types
(\texttt{regtypes}), Hindley-Milner (\texttt{hmtypes}), etc., are
provided as libraries. Since, e.g., \texttt{list\_of/2} in
Fig.~\ref{fig:fmoreprops} is a property that is in addition a regular
type, this can be flagged as \codehlt{:- prop list\_of/2 + regtype.}
or, more compactly, \codehlt{:- regtype list\_of/2.}
Most properties (including types) are ``runnable'' (useful for
run-time checking), and can be interacted with, i.e., the answers to a
query \texttt{?- use\_package(someprops), X = \~{}list.} are:
\texttt{X = []}, \texttt{X = [\_]}, \texttt{X = [\_,\_]}, \texttt{X =
  [\_,\_,\_]}, etc.
Note also that assertions such as the one in line~\ref{listuse} 
provide information not only on (a generalization of) types 
but also on modes. 

\begin{figure}[t]
\begin{lstlisting}
:- module(_, [nrev/2], [assertions, nativeprops, functional]).
:- entry nrev/2 : {list, ground} * var. (*@\label{entry}@*)
:- use_module(someprops).

:- pred nrev(A, B) : list(A) => list(B). (*@\label{listuse}@*)
:- pred nrev(A, B) : list_of(color, A) => list_of(color, B). (*@\label{colorlistuse}@*)
:- pred nrev(A, B) : list(A) + (not_fails, is_det, terminates). (*@\label{conj}@*)
:- pred nrev(A, _) : list(A) + steps_o(length(A)). (*@\label{linear}@*)

nrev([])    := [].
nrev([H|L]) := ~conc(nrev(L),[H]).

:- pred conc(A,B,C) : list(A) => size_ub(C,length(A)+length(B)) 
                              +  steps_o(length(A)).
conc([],    L) := L.
conc([H|L], K) := [ H | conc(L,K) ]. 
\end{lstlisting}
\figskip
\caption{Naive reverse with some --partially erroneous-- assertions.}
\label{fig:nrevf}
\underfig
\end{figure}

In general \texttt{pred} assertions follow the schema:

\smallskip\noindent
\codehlt{\kbd{:- pred} \nt{Pred} [\kbd{:} \nt{Precond\/}]
[\kbd{=>} \nt{Postcond\/}] [\kbd{+} \nt{CompProps\/}]\kbd{.}}

\smallskip\noindent
\nt{Pred} is a {\em predicate descriptor}, i.e., a predicate symbol
applied to distinct free variables, such as, e.g.,
\texttt{nrev(A,B)}. \nt{Precond} and \nt{Postcond} are logic formulas
about execution states, that we call \nt{StateFormulas}. An execution
state is defined by the bindings of values to variables in a given
execution step (in logic programming terminology, a substitution, plus
any global state).  An atomic \nt{StateFormula} 
(such as, e.g., \texttt{list(X)}, \texttt{X > 3}, or
\texttt{sorted(X)}) is a literal whose 
predicate symbol corresponds to a state property. 
A \nt{StateFormula} can also be a
conjunction or disjunction of \nt{StateFormulas}.  Standard (C)LP
syntax is used, with comma representing conjunction (e.g.,
``\texttt{(list(X), list(Y))}'') and semicolon disjunction (e.g.,
``\texttt{(list(X) ; int(X))}'').  
\nt{Precond} is the precondition under which the
\texttt{pred} assertion is applicable.
\nt{Postcond} states a conditional postcondition, i.e., it expresses
that in any call to \nt{Pred}, if \nt{Precond} holds in the calling
state and the computation of the call succeeds, then \nt{Postcond}
should also succeed in the success state.
If \nt{Precond} is omitted, the assertion is equivalent to: 
\ 
\codehlt{\kbd{:- pred} \nt{Pred} \kbd{:} \kbd{true} \kbd{=>} \nt{Postcond}\kbd{.}} 
and it is interpreted as ``for any call to \nt{Pred} which succeeds,
\nt{Postcond} should succeed in the success state.''
As Fig.~\ref{fig:nrevf} shows, there can be
several \texttt{pred} assertions for the same predicate. The set of
preconditions (\nt{Precond}) in those assertions is considered {\em
  closed} in the sense that they must cover all valid calls to the
predicate.

Finally, \texttt{pred} assertions can include a
\nt{CompProps} field, used to describe properties of the whole
computation of the calls to predicate \nt{Pred} that meet precondition
\nt{Precond}. For example, the assertion
in line~\ref{linear} of Fig.~\ref{fig:nrevf},
states that for any call to predicate \texttt{nrev/2} with the first
argument bound to a list, the number of resolution steps, given as a
function on the length of list $A$, is in $O(length(A))$ (i.e., such
function is linear in $length(A)$).\footnote{This is of course false,
  but we will let the compiler tell us --see later.}
The assertion in line~\ref{conj} of Fig.~\ref{fig:nrevf} is an example
where \nt{CompProps} is a conjunction:
it expresses that the previous calls do not fail without first
producing at least one solution, are deterministic (i.e., they produce
at most one solution at most once), and terminate.  Thus, in this
case, \nt{CompProps} describes a terminating functional computation.
The rest of the assertions in Fig.~\ref{fig:nrevf} will be explained
later, in the appropriate sections.

In order to facilitate writing assertions, \ciao\ also provides
additional syntactic sugar such as \emph{modes} and cartesian product
notation.
For example, consider the following set of \texttt{pred} assertions
providing information on a reversible sorting predicate:

\smallskip\noindent
\codehltw{%
:- pred sort/2 : list(num) * var => list(num) * list(num) + is\_det. \\
:- pred sort/2 : var * list(num) => list(num) * list(num) + non\_det. }

\smallskip\noindent
(in addition, curly brackets can be used to group properties --see
Fig.~\ref{fig:qsort}). 
Using \ciao's \texttt{isomodes} library, which provides syntax and
meaning for the ISO instantiation operators, this can also be expressed
as:

\smallskip\noindent
\codehltw{:- pred sort(+list(num), -list(num)) + is\_det.\\
:- pred sort(-list(num), +list(num)) + non\_det.}

\smallskip
The \texttt{pred} assertion schema is in fact syntactic
sugar for
combinations of atomic assertions of the following three types:

\smallskip\noindent
\codehltw{%
\kbd{:- calls\ \ } \nt{Pred} [\kbd{:} \nt{Precond\/}]\kbd{.} \\
\kbd{:- success} \nt{Pred} [\kbd{:} \nt{Precond\/}] [\kbd{=>} \nt{Postcond\/}]\kbd{.} \\
\kbd{:- comp\ \ \ } \nt{Pred} [\kbd{:} \nt{Precond\/}] [\kbd{+} \nt{CompProps\/}]\kbd{.}
} \\

\smallskip\noindent
which describe all the admissible call states, the success states, and computational
properties for each set of admissible call states (in this order).

\noindent
\textbf{Program-point assertions} are of the form
\codehlt{check(\nt{StateFormula}\ )} and they can be placed at the
locations in programs in which a new literal may be added.  They
should be interpreted as ``whenever computation reaches a state
corresponding to the program point in which the assertion is,
\nt{StateFormula} should hold.'' For example,

\smallskip\noindent
\codehlt{check((list\_of(color, A), var(B)))} 

\smallskip\noindent is a program-point assertion, where \texttt{A} and
\texttt{B} are variables of the clause where the assertion appears.

\parbegin
\paragraph{\textbf{Assertion status:}}
Independently of the schema, each assertion
can be in a \emph{verification status},
marked by prefixing the assertion itself with the keywords,
\texttt{check}, \texttt{trust}, \texttt{true}, \texttt{checked}, and
\texttt{false}.  This specifies respectively whether the assertion is
provided by the programmer and is to be checked or to be trusted, or
is the output of static analysis and thus correct (safely
approximated) information, or the result of processing an input
assertion and proving it correct or false, as will be discussed in the
next section.  The \texttt{check} status is assumed by default when no
explicit status keyword is present (as in the examples so far).

\parbegin
\paragraph{\textbf{Uses of assertions:}} 
as we will see, assertions find many uses in \ciao, ranging from
testing to verification and documentation (for the latter, see
\texttt{lpdoc}~\cite{lpdoc-cl2000-short}).  In addition to describing
the properties of the module in which they appear, assertions also
allow programmers to describe properties of modules / classes which
are not yet written or are written in other languages.\footnote{This
  is also done in other languages but, in contrast with \ciao,
  different kinds of assertions for each purpose are often used.}
This makes it possible to run checkers / verifiers / documenters
against partially developed code.

\section{The \ciao\  Unified  Assertion  Framework}
\label{sec:assert-philosophy}

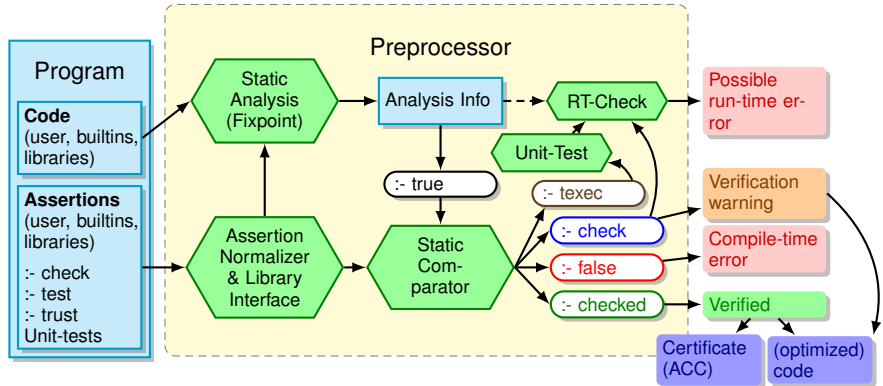
\begin{figure}[t]
  \centering
\pgfdeclarelayer{background}
\pgfdeclarelayer{foreground}
\pgfsetlayers{background,main,foreground}

\tikzstyle{source}=[draw, draw=cyan!80!black!100, fill=cyan!20, text width=6em, font=\sffamily,
    thick,
    minimum height=2.5em,drop shadow]
\tikzstyle{tool}=[draw=green!50!black!100, fill=green!40, text width=5em, font=\sffamily, 
    thick,
    text centered, 
    chamfered rectangle, chamfered rectangle angle=30, chamfered rectangle xsep=2cm]
\tikzstyle{midresult}=[draw, fill=white!40, text width=5em, font=\sffamily,
    thick,
    rounded rectangle,
    minimum height=1em,drop shadow]
\tikzstyle{warnresult}=[color=orange!50!black!100, fill=orange!40, text width=6em, font=\sffamily, 
    thick,
    rounded corners=2pt,
    minimum height=1em,drop shadow]
\tikzstyle{errresult}=[color=red!80!black!100, fill=red!20, text width=6em, font=\sffamily, 
    thick,
    rounded corners=2pt,
    minimum height=1em,drop shadow]
\tikzstyle{okresult}=[color=green!50!black!100, fill=green!40, text width=6em, font=\sffamily, 
    thick,
    rounded corners=2pt,
    minimum height=1em,drop shadow]
\tikzstyle{certresult}=[color=blue!50!black!100, fill=blue!40, text width=5em, font=\sffamily, 
    thick,
    rounded corners=2pt,
    minimum height=1em,drop shadow]
\tikzstyle{coderesult}=[color=blue!50!black!100, fill=blue!40, text width=5em, font=\sffamily, 
    thick,
    rounded corners=2pt,
    minimum height=1em,drop shadow]

\scriptsize
\begin{tikzpicture}[>=latex]
  \node (code) [source] {
    \textbf{Code}\\
    (\mbox{user, builtins,} libraries)
  };
  \path (code)+(0,-7em) node (assertions) [source] {
    \textbf{Assertions}\\
    (\mbox{user, builtins,} libraries)\\[1ex]
    :- check\\
    :- test\\
    :- trust\\
    Unit-tests \\
  };

\path (code)+(10em,2em) node (statana) [tool] {Static Analysis (Fixpoint)};
\path (statana)+(0,-9em) node (normalizer) [tool] {Assertion Normalizer \& Library Interface};
\path (statana)+(9.5em,0) node (anainfo) [source] {Analysis Info};
\path[color=black] (anainfo)+(0em,-4.5em) node (true) [midresult] {:- true};
\path (anainfo)+(0em,-9em) node (comparator) [tool] {Static Comparator};
\path (anainfo)+(9em,0) node (rtcheck) [tool] {RT-Check};
\path (rtcheck)+(-3em,-2.8em) node (unittest) [tool] {Unit-Test};
\path[color=brown!50!black] (comparator)+(8em,4em) node (texec) [midresult] {:- texec};
\path[color=blue] (comparator)+(9em,2em) node (check) [midresult] {:- check};
\path[color=red] (comparator)+(9em,0em) node (false) [midresult] {:- false};
\path[color=green!50!black] (comparator)+(9em,-2em) node (checked) [midresult] {:- checked};

\path (rtcheck)+(8.5em,0) node (rterror) [errresult] {Possible\\run-time error};
\path (false)+(8.5em,1em) node (cterror) [errresult] {Compile-time error};
\path (check)+(8.5em,2em) node (verifwarn) [warnresult] {Verification warning};
\path (checked)+(8.5em,0) node (verified) [okresult] {Verified};
\path (verified)+(-3em,-3em) node (certificate) [certresult] {Certificate (ACC)};
\path (verified)+(3em,-3em) node (optcode) [coderesult] {(optimized) code};

\path [draw, thick, ->] (code.east) -- node [] {} (statana.west) ;
\path [draw, thick, ->] (assertions) -- node [] {} (normalizer.west) ;
\path [draw, thick, ->] (normalizer) -- node [] {} (statana) ;
\path [draw, thick, ->] (normalizer.east) -- node [] {} (comparator.west) ;
\path [draw, thick, ->] (comparator.east) -- node [] {} (texec.south west) ;
\path [draw, thick, ->] (comparator.east) -- node [] {} (check.west) ;
\path [draw, thick, ->] (statana) -- node [] {} (anainfo) ;
\path [draw, thick, ->] (comparator.east) -- node [] {} (false.west) ;
\path [draw, thick, ->] (comparator.east) -- node [] {} (checked.west) ;
\path [draw, thick, ->] (anainfo) -- node [] {} (true) ;
\path [draw, thick, ->] (true) -- node [] {} (comparator) ;
\path [draw, densely dashed, thick, ->] (anainfo) -- node [] {} (rtcheck) ;
\draw [thick, ->] (texec.north east) to [bend right] (unittest.south east) ;
\path [draw, thick, ->] (unittest) -- node [] {} (rtcheck) ;
\path [draw, thick, ->] (rtcheck) -- node [] {} (rterror) ;
\draw [thick, ->] (check.north east) to [bend right] (rtcheck) ;
\path [draw, thick, ->] (check) -- node [] {} (verifwarn) ;
\path [draw, thick, ->] (false) -- node [] {} (cterror) ;
\path [draw, thick, ->] (checked) -- node [] {} (verified) ;
\path [draw, thick, ->] (verified) -- node [] {} (certificate) ;
\path [draw, thick, ->] (verified) -- node [] {} (optcode) ;
\draw [thick, ->] (verifwarn.east) to [bend left] (optcode.north east) ;

\path [color=black] (anainfo.north)+(0,1.5em) node (preprocessor) {\sffamily \small Preprocessor};

\path [color=black] (code.north)+(0,1.5em) node (program) {\sffamily \small Program};

\begin{pgfonlayer}{background}
  \path (statana.north west)+(-0.5,1.0) node (g) {};
  \path (checked.south east)+(0.5,-0.5) node (h) {};
  
  \path[fill=yellow!20,rounded corners, draw=black!50, densely dashed] (g) rectangle (h);
\end{pgfonlayer}

\begin{pgfonlayer}{background}
  \path (code.north west)+(-0.1,0.8) node (g) {};
  \path (assertions.south east)+(0.1,-0.1) node (h) {};
  
  \path[source] (g) rectangle (h);
\end{pgfonlayer}

\end{tikzpicture}

  \figskip
  \caption{The \ciao\ assertion framework (\ciaopp's
    verification/testing architecture).} 
  \label{fig:integrated-framework-ciaopp}
\underfig
\end{figure}

We now describe the \ciao\ unified assertion
framework~\cite{aadebug97-informal-short,prog-glob-an-short,assert-lang-disciplbook-short},
implemented in the \ciao\ preprocessor, \ciaopp.
Fig.~\ref{fig:integrated-framework-ciaopp} depicts the overall
architecture. Hexagons represent tools and arrows indicate the
communication paths among them.  It is a design objective of the
framework that most of this communication be performed also in terms
of assertions. This has the advantage that at any point in the process
the information is easily readable by the user.
The input to the process is the user program, \emph{optionally}
including a set of assertions; this set always includes any 
assertion present for predicates exported by 
any libraries used (left part of
Fig.~\ref{fig:integrated-framework-ciaopp}).

\noindent
\textbf{Run-time checking of assertions:} after (assertion)
normalization (which, e.g., takes away syntactic sugar) the {\em
  RT-check} module transforms the program by adding run-time checks to
it that encode the meaning of the assertions (we assume for now that
the {\em Comparator} simply passes the assertions through). Note that
the fact that properties are written in the source language and
\emph{runnable}
is very useful in this process. Failure of these checks raises
run-time errors referring to the corresponding assertion.
{\em Correctness} of the transformation requires that the transformed
program only produce an error if the assertion is in fact
violated. 

\noindent
\textbf{Compile-time checking of assertions:} even though run-time
checking can detect violations of specifications, it cannot guarantee
that an assertion holds. Also, it introduces run-time overhead.  The
framework performs compile-time checking of assertions by
\emph{comparing} the results of \emph{Static Analysis}
(Fig.~\ref{fig:integrated-framework-ciaopp}) with the
assertions~\cite{aadebug97-informal-short,prog-glob-an-short}. This
analysis is typically performed by abstract
interpretation~\cite{Cousot77-short} or any other mechanism that
provides \emph{safe} upper or lower approximations of relevant
properties, so that comparison with assertions is meaningful despite
precision losses in the analysis.  The type of analysis may be
selected by the user or determined automatically based on the
properties appearing in the assertions.  Analysis results are given
using also the assertion language, to ensure interoperability and make
them understandable by the programmer.  As a possible result of the
comparison, assertions may be proved to hold, in which case they get
\texttt{checked} status --Fig.~\ref{fig:integrated-framework-ciaopp}.
If all assertions are \texttt{checked} then the program is verified.
In that case a certificate can be generated that can be shipped with
programs and checked easily at the receiving end (using the
\emph{abstraction carrying code}
approach~\cite{ai-safety-ngc07-short}).
As another possible result, assertions can be proved not to hold, in
which case they get \texttt{false} status and a \emph{compile-time
  error} is reported.  Even if a program contains no assertions, it
can be checked against the assertions contained in the libraries used
by the program, potentially catching bugs at compile time.
Finally, and most importantly, if it is not possible to prove nor to
disprove (part of) an assertion, then such assertion (or part) is left
as a \texttt{check} assertion, for which optionally run-time checks
can be generated as described above. This can optionally produce a
\emph{verification warning}.

The fact that the system deals throughout with safe approximations of
the meaning of the program, and that remaining in \texttt{check}
status is an acceptable outcome of the comparison process, allows
dealing with complex properties in a correct way.  For example, in
\ciaopp\ the programmer has the possibility of stating assertions
about the efficiency of the program (lower and/or upper bounds on the
computational cost of procedures~\cite{resource-verif-iclp2010-short})
which the system will try to verify or falsify, thus performing
automatic debugging and verification of the \emph{performance} of
programs (see Section~\ref{sec:static-verif-debug}).
Other interesting properties are handled such as data structure shape
(including pointer sharing), bounds on data structure sizes, and other
operational properties, as well as procedure-level properties such as
determinacy~\cite{determinacy-ngc09-short},
non-failure~\cite{nfplai-flops04-short}, termination, and bounds on
the execution time~\cite{estim-exec-time-ppdp08-short}, and the
consumption of a large class of user-defined
resources~\cite{resource-iclp07-short}.
Assertion checking in \ciaopp\ is also
module-aware~\cite{mod-ctchecks-lpar06-short,mod-types-pepm08-short}.
Finally, the information from analysis can be used to optimize the
program in later compilation stages, as we will discuss later.

\begin{figure}[t]
\reducespace
\begin{lstlisting}
:- module(qsort, [qsort/2], [assertions, functional]).
:- use_module(compare, [geq/2, lt/2]).
:- entry qsort/2 : {list(num), ground} * var.
 
qsort([])    := [].
qsort([X|L]) := ~conc(qsort(L1), [X|qsort(L2)]) 
             :- partition(L, X, L1, L2).
 
partition([],_B,[],[]).
partition([E|R],C,[E|Left1],Right) :- 
        lt(E,C),  partition(R,C,Left1,Right).
partition([E|R],C,Left,[E|Right1]) :-
        geq(E,C), partition(R,C,Left,Right1).
\end{lstlisting}
\figskip
  \caption{A modular \kbd{qsort} program.}
  \label{fig:qsort}
\underfig
\end{figure}
 
\section{Static Verification, Debugging, Run-Time Checking, and Unit
  Testing in Practice}
\label{sec:progr-docum-stat}

We now present some examples which illustrate the use of the \ciao\
assertion framework discussed in the previous section, as implemented
in \ciaopp.  We also introduce some more examples of the assertion
language as we proceed.

\subsection{Automatic Inference of (Non-Trivial) Code Properties}

We first illustrate with examples the automatic inference of
code properties (box ``Static Analysis'' in
Fig.~\ref{fig:integrated-framework-ciaopp}). 
Modes and types are inferred, as mentioned before, using different
methods including ~\cite{ai-jlp-short,freeness-iclp91-short} for modes
and~\cite{Saglam-Gallagher-95-short,eterms-sas02-short} for types.
As also mentioned before, \ciaopp\ includes a non-failure
analysis~\cite{nfplai-flops04-short}, which can detect procedures and
goals that can be guaranteed not to fail, i.e., to produce at least
one solution or not to terminate. It also can detect predicates that
are ``covered'', i.e., such that for any input (included in the
calling type of the predicate), there is at least one clause whose
``test'' (head unification and body builtins) succeeds.  \ciaopp\ also
includes a determinacy analysis~\cite{determinacy-ngc09-short}, which
can detect predicates which produce at most one solution at most once,
or predicates whose clause tests are mutually exclusive, even if they
are not deterministic because they call other predicates that can
produce more than one solution (it means that the predicate does not
perform backtracking at the level of its clauses).

Consider again the naive reverse program in Fig.~\ref{fig:nrevf}.  The
assertion in line~\ref{entry} is an example of an \texttt{entry}
assertion: a \texttt{pred} assertion addressing calls from outside the
module.\footnote{Note that in \ciaopp\ the \texttt{pred} assertions of
  exported predicates can be used optionally instead of
  \texttt{entry}.}  It informs the \ciaopp\ analyzers that in all
external calls to \texttt{nrev/2}, the first argument will be a ground
list and the second one a free variable. Using only the information
specified in the \texttt{entry} assertion, the aforementioned analyses
infer different sorts of information which include, among others, that
expressed by the following assertion:

\negskip
\begin{small}
\begin{verbatim}
:- true pred nrev(A,B): ( list(A), var(B) ) => ( list(A), list(B) )
                      + ( not_fails, covered, is_det, mut_exclusive ).
\end{verbatim}
\negskip
\end{small}

As mentioned before, \ciaopp\ can also infer lower and upper bounds on
the sizes of terms and the computational cost of
predicates~\cite{low-bounds-ilps97-short,granularity-short,caslog-short},
including user-defined resources~\cite{resource-iclp07-short}.  The
cost bounds are expressed as functions on the sizes of the input
arguments and yield the number of resolution steps.
Note that obtaining a finite upper bound on cost also implies proving
{\em termination} of the predicate.

As an example, the following assertion is part of the output of the
lower-bounds analysis (that also includes a non-failure analysis,
without which a trivial lower bound of 0 would be derived):

\negskip
\begin{small}
\begin{verbatim}
:- true pred conc(A,B,C) : ( list(A), list(B), var(C) )
                        => ( list(A), list(B), list(C),
                             size_lb(A,length(A)), size_lb(B,length(B)),
                             size_lb(C,length(B)+length(A)) )
                         + ( not_fails, covered, steps_lb(length(A)+1)).
\end{verbatim}
\end{small}
\negskip Note that in this example the size measure used is list
length.  The
property \\
{\small \verb-size_lb(C,length(B)+length(A))-} means that a (lower)
bound on the size of the third argument of \texttt{conc/3} is the sum
of the sizes of the first and second arguments.  The inferred lower
bound on computational steps is the length of the first argument of
\texttt{conc/3} plus one.
The \texttt{length/1} property used in the previous assertion is just
the \texttt{length/2} predicate called using functional syntax, that
curries the last argument.  \ciaopp~currently uses some predefined
metrics for measuring the ``size'' of an input, such as list length,
term size, term depth, or integer value.  These are automatically
assigned to the predicate arguments involved in the size and cost
analysis according to the previously inferred type information.  A
new, experimental version of the size analyzers is in development that
can deal with user-defined size metrics (i.e., predicates) and is also
able to synthesize automatically size metrics.

\subsection{Static (Performance) Verification and Debugging}
\label{sec:static-verif-debug}
We now illustrate static verification and debugging, i.e., statically
proving or disproving program assertions (i.e., specifications).  This
corresponds to the ``Static Comparator'' box in
Fig.~\ref{fig:integrated-framework-ciaopp}.  We focus on verification
of the resource usage of programs, such as lower and/or upper bounds
on execution steps or user defined resources, but the process also
applies to more traditional properties such as types and modes.
Consider the assertion in line~\ref{linear} of Fig.~\ref{fig:nrevf},
which states that \texttt{nrev} should be linear in the length of the
(input) argument \texttt{A}.
With compile-time error checking turned on, \ciaopp\ automatically
selects mode, type, non-failure, and lower/upper-bound cost analyses and
issues the following error message (corresponding to the ``compile-time error'' exit
in Fig.~\ref{fig:integrated-framework-ciaopp}):

\negskip
\begin{small}
\begin{verbatim}
ERROR: False assertion:
            :- pred nrev(A, _) : list(A) + steps_o(length(A))
       because on comp nrev:nrev(A,_):
       [generic_comp] : steps_lb(0.5*exp(length(A),2)+1.5*length(A)+1)
\end{verbatim}
\end{small}
\negskip

\noindent
This message states that \kbd{nrev} will take at least
$\frac{length(A)^2 + 3 \ length(A)}{2} + 1$ resolution steps (a safe
lower bound inferred by the cost analyzer), while the assertion
requires the cost to be in $O(length(A))$ resolution steps.  As a
result, the worst case asymptotic complexity stated in the
user-provided assertion is proved wrong by the lower bound cost
assertion inferred by the analysis.  Note that upper-bound cost
assertions can be proved to hold by means of upper-bound cost analysis
if the bound computed by analysis is lower or equal than the upper
bound stated by the user in the assertion. The converse holds for
lower-bound cost
assertions~\cite{aadebug97-informal-short,resource-verif-iclp2010-short}.
Thanks to this functionality, \ciaopp\ can also certify programs with
resource consumption assurances as well as efficiently checking such
certificates~\cite{tutorial-europar04-short}.

\subsection{Run-Time Checking}

\begin{figure}[t]
\reducespace
\begin{lstlisting}[firstnumber=5]
:- pred qsort(A,B) => (ground(B),sorted_num_list(B)).

:- prop sorted_num_list/1.

sorted_num_list([]).
sorted_num_list([X]):- num(X).
sorted_num_list([X,Y|Z]):- num(X),num(Y),geq(Y,X),
                           sorted_num_list([Y|Z]).
qsort([],[]).
qsort([X|L],R) :- partition(L,X,L1,L2), 
                  qsort(L2,R2), qsort(L1,R1), 
                  conc(R2,[X|R1],R).
\end{lstlisting}
\figskip
\caption{An example for run-time checking.}
\label{fig:rttests}
\underfig
\end{figure}

As mentioned before, (parts of) assertions which cannot be verified at
compile time (see again Fig.~\ref{fig:integrated-framework-ciaopp})
are translated into run-time checks via a program transformation. 
As an example, consider the assertion, property definitions, and
(wrong) definition of \texttt{qsort/2} in Fig.~\ref{fig:rttests}
(where \texttt{partition/4} and \texttt{conc/3} are defined as in
Figures~\ref{fig:qsort} and~\ref{fig:fnrev} respectively).  The
assertion states that \texttt{qsort/2} always returns a ground,
\emph{sorted} list of numbers. The program contains a bug to be
discovered.
With run-time checking turned on, the following query produces the
listed results:

\negskip
\begin{small}
\begin{verbatim}
?- qsort([1,2],X).
{In /tmp/qsort.pl
ERROR: (lns 5-5) Run-time check failure in assertion for: qsort:qsort/2.
In *success*, unsatisfied property: sorted_num_list.
ERROR: (lns 13-16) Failed in qsort:qsort/2.}
\end{verbatim}
\end{small}
\negskip

Two errors are reported for a single run-time check failure: the first
error shows the actual assertion being violated and the second marks
the first clause of the predicate which violates the assertion.
However, not enough information is provided to determine which literal
made the erroneous call.  
It is also possible to increase the verbosity level of the messages
and to produce a call stack dump up to the exact program point where
the violation occurs, showing for each predicate the body literal that
led to the violation:

\negskip

\begin{small}
\begin{verbatim}
?- set_ciao_flag(rtchecks_callloc,literal),
   set_ciao_flag(rtchecks_namefmt,long), use_module('/tmp/qsort.pl').
yes
?- qsort([3,1,2],X).
{In /tmp/qsort.pl
ERROR: (lns 5-5) Run-time check failure in assertion for: qsort:qsort(A,B).
In *success*, unsatisfied property: sorted_num_list(B).
Because: ['B'=[2,1]].
ERROR: (lns 13-16) Failed in qsort:qsort(A,B).
ERROR: (lns 13-16) Failed when invocation of qsort:qsort([X|L],R)
       called qsort:qsort(L1,R1) in its body.}
{In /tmp/qsort.pl
ERROR: (lns 5-5) Run-time check failure in assertion for: qsort:qsort(A,B).
In *success*, unsatisfied property: sorted_num_list(B).
Because: ['B'=[3,2,1]].
ERROR: (lns 13-16) Failed in qsort:qsort(A,B).}
\end{verbatim}
\end{small}
\negskip

The output makes it easier to locate the error since 
the call stack dump provides the list of calling predicates.
Note that the first part of the assertion is not violated, since
\texttt{B} is ground.  However, on success the output of
\texttt{qsort/2} is a sorted list but in reverse order, which gives us
a hint: the variables \texttt{R1} and \texttt{R2} in the call to
\texttt{conc/3} are swapped by mistake.

\subsection{Unit Testing} 
\label{sec:unit-testing}

Unit tests need to express on one hand {\em what to execute} and on
the other hand {\em what to check} (at run time). A key characteristic
of the \ciao\ approach to unit testing
(see~\cite{testchecks-iclp09-short} for a full description) is that it
(re)uses the assertion language for expressing what to check. This
avoids redundancies and allows reusing the same assertions and
properties used for static and/or run-time checking.  However, the
assertion language does include a minimal number of additional
elements for expressing {\em what to execute}.  In particular, it
includes the following assertion schema:
\smallskip\noindent
\codehlt{\kbd{:- texec} \nt{Pred} [\kbd{:} \nt{Precond}]
[\kbd{+} \nt{ExecProps\/}]\kbd{.}}

\smallskip\noindent which states that we want to execute (as a test) a
call to \nt{Pred} with its variables instantiated to values that
satisfy \nt{Precond}.  \nt{ExecProps} is a conjunction of properties
describing how to drive this execution.  As an example, the assertion:

\smallskip\noindent
\codehlt{:- texec conc(A, B, C) : (A=[1,2],B=[3],var(C)).} 

\smallskip\noindent
expresses that the testing harness should execute a call to
\texttt{conc/3} with the first and second arguments bound to
\texttt{[1,2]} and \texttt{[3]} respectively and the third one
unbound.

In our approach many of the properties that can be 
used in \nt{Precond} (e.g., types) can also be used as value generators 
for those variables, so that input data  can be automatically generated for the unit
tests (see e.g., the technique described in~\cite{tdg-prolog-wlpe08}). 
However, there are also some properties that are specific for this
purpose, such as, e.g., random value generators.

We can define a complete unit test using the \texttt{texec} assertion
together with other assertions expressing {\em what to
  check at run time} such as, for example: 

\smallskip\noindent
\codehltw{%
:- check success conc(A,B,C):(A=[1,2],B=[3],var(C)) => C=[1,2,3]. \\
:- check comp \ \ \ conc(A,B,C):(A=[1,2],B=[3],var(C)) + not\_fails.
} 

\smallskip
The success assertion states that if a call to \texttt{conc/3} with
the first and second arguments bound to \texttt{[1,2]} and
\texttt{[3]} respectively and the third one unbound terminates with
success, then the third argument should be bound to
\texttt{[1,2,3]}. The comp assertion says that such a call should not
fail.

One additional advantage of \ciao's unified framework is that the
execution expressed by a 
\nt{Precond} in a \texttt{texec} assertion for unit testing can also
trigger the checking of parts of other assertions that could not be
checked at compile time and thus remain as run-time checks. This way,
a single set of run-time checking ma\-chi\-ne\-ry can deal with both
run-time checks and unit tests.  Conversely, static checking of
assertions can safely avoid (possibly parts of) unit test execution
(see Fig.~\ref{fig:integrated-framework-ciaopp} again), so that
sometimes unit tests can be checked without ever running them.

Finally, the system provides as syntactic sugar another predicate
assertion schema, the {\tt test} schema:
\smallskip\noindent
\codehlt{\kbd{:- test} \nt{Pred} [\kbd{:} \nt{Precond\/}]
[\kbd{=>} \nt{Postcond\/}] [\kbd{+} \nt{CompExecProps\/}]\kbd{.}}

\smallskip\noindent
This assertion is interpreted as the combination of the following three
assertions: 

\smallskip\noindent
\codehltw{%
\kbd{:- \ \ \ \ \ \ texec} \ \ \nt{Pred} [\kbd{:} \nt{Precond\/}] [\kbd{+} \nt{ExecProps\/}]\kbd{.}\\
\kbd{:- check success} \nt{Pred} [\kbd{:} \nt{Precond\/}]
  [\kbd{=>} \nt{Postcond\/}]\kbd{.}\\
\kbd{:- check comp\ \ \ } \nt{Pred} [\kbd{:} \nt{Precond\/}]
  [\kbd{+} \nt{CompProps\/}]\kbd{.}
}

\smallskip\noindent
For example, the assertion: 

\smallskip\noindent
\codehlt{%
\kbd{:- test} \texttt{conc(A,B,C)\kbd{:}
  (A=[1,2],B=[3],var(C))\kbd{=>} C=[1,2,3] \kbd{+} not\_fails\kbd{.}}}

\smallskip\noindent
is conceptually equivalent to the three (texec, success, comp) 
shown previously as examples (\nt{CompExecProps} being the
conjunction of \nt{ExecProps} and \nt{CompProps}). 

The assertion language not only allows checking single solutions (as
it is done in the previous \texttt{test} assertion for
\texttt{conc}/3), but also multiple solutions to calls. 
In addition, it includes a set of predefined properties that can be
used in \nt{ExecProps} that are specially useful in the context of
unit tests, including: an upper bound \texttt{N} on the number of
solutions to be checked (\texttt{try\_sols(N)}); expressing that the
execution of the unit test should be repeated \texttt{N} times
(\texttt{times(N)}); that a test execution should throw a particular
exception (\texttt{exception(Excep)}); or that a predicate should
write a given string into the current output stream
(\texttt{user\_output(String)}) or the current error stream
(\texttt{user\_error(String)}).
Similarly, properties are provided that are useful in \nt{Precond},
for example, to generate random input data with a given probability
distribution (e.g., for floating point numbers, including special
cases like {\em infinite}, {\em not-a-number}, or {\em zero} with
sign).

The testing mechanism has proved very useful in practice. For example,
with it we have developed a battery of tests that are used for
checking ISO-Prolog compliance in \ciao. The set contains 976 unit
tests, based on the \emph{Stdprolog}
application~\cite{DBLP:conf/iclp/SzaboS06}.

\section{High Performance with Less Effort}
\label{sec:very-high-perf}

A potential benefit of strongly typed languages is performance: the
compiler can generate more efficient code with the additional type and
mode information that the user provides.  Performance is a good thing,
of course. However, it is also attractive to avoid putting the burden
of efficient compilation on the user by requiring the presence of many
program declarations: the compiler should certainly take advantage of
any information given by the user, but if the information is not
available, it should do the work of inferring such program properties
whenever possible.
This is the approach taken in \ciao: as we have seen before,
when assertions are not present in the program, \ciao's analyzers try
to \emph{infer} them.  Most of these analyses are performed at the
kernel language level, so that the same analyzers are used for several
of the supported programming models.

\parbegin
\paragraph{\textbf{High-level optimization:}}
the information inferred by the global analyzers is used to perform
high-level optimizations, including multiple abstract
specialization~\cite{spec-pepm-short}, partial
evaluation~\cite{ai-with-specs-sas06-short}, dead code removal, goal
reordering, reduction of concurrency / dynamic
scheduling~\cite{spec-dyn-sch-iclp97-short}, etc.

\parbegin
\paragraph{\textbf{Optimizing compilation:}}
the objective is again to achieve the best of both worlds: with no
assertions or analysis information, the low-level \ciao\ compiler
(\texttt{ciaoc}~\cite{ciaoc-entcs-short}) generates code which is
competitive in speed and size with the best dynamically typed systems.
And then, when useful information is present, either coming from the
user or inferred by the system analyzers, the experimental optimizing
compiler, \texttt{optimcomp} (see,
e.g.,~\cite{morales04:p-to-c-padl-short} for an early description) can
produce code that is competitive with that of strongly-typed systems.
\ciao's highly optimized compilation has been successfully tested, for
example, in applications with tight memory and real-time
constraints~\cite{carro06:stream_interpreter_cases}, obtaining a
7-fold speed-up w.r.t.\ the default bytecode compilation.  The
performance of the latter is already similar to that of
state-of-the-art abstract machine-based systems.  The application
involved the real-time spatial placement of sound sources for a
virtual reality suit, and ran in a small (``Gumstix'') processor
embedded within a headset.  Interestingly, this performance level is
only around 20-40\% slower than a comparable (but more involved)
implementation in C of the same application.

\parbegin
\paragraph{\textbf{ImProlog:}} driven by the need of producing
efficient final code in extreme cases, we have also introduced in the
more experimental parts of the system the design and compilation of a
variant of Prolog (which we termed \emph{ImProlog}) which, besides
assertions for types and modes, introduces imperative features such as
\emph{low-level pointers} and \emph{destructive assignment}.  This
restricted subset of the merge of the imperative and logic paradigms
is present (in beta) in the \texttt{optimcomp} branch and has been
used to write a complete WAM emulator including its
instructions~\cite{morales09:improlog-tr}, and part of its lower-level
data structures~\cite{tagschemes-ppdp08-short}.  This source code is
subject to several analysis and optimization stages to generate highly
efficient C code.
This approach is backed by some early performance numbers, which show
this automatically generated machine to be on average just 8\% slower
than that of a highly optimized emulator, such as
\yap~5.1.2~\cite{yap02} (and actually faster in some benchmarks), and
44\% faster than the stock \ciao\ emulator.
In this case, some of the annotations ImProlog takes advantage of
cannot be inferred by the analyzers because, for example, they address
issues (such as word size) which depend on the targeted architecture,
which must be entered by hand. 

\parbegin
\paragraph{\textbf{Automatic parallelization:}}
a particularly interesting optimization performed by \ciaopp, in the
same vein of obtaining high performance with less effort from the
programmer, and which
is inherited from the \&-Prolog system, is \emph{automatic
  parallelization}~\cite{tutorial-europar97-short,partut-toplas-short}.
This is specially relevant nowadays given that the wide availability of
multicore processors has made parallel computers mainstream.
We illustrate this by means of a simple example using goal-level
program
parallelization~\cite{effofai-toplas-short,uudg-annotators-lopstr2007-short}.
This optimization is performed as a source-to-source transformation,
in which the input program is \emph{annotated} with parallel
expressions as a result.  The parallelization algorithms, or
annotators~\cite{annotators-jlp-short}, exploit parallelism under
certain \emph{independence} conditions, which allow guaranteeing
interesting correctness and no-slowdown properties for the
parallelized programs~\cite{sinsi-jlp-short,consind-toplas-short}.
This process is made more complex by the presence of variables shared
among goals and pointers among data structures at run time.

\begin{figure}[t]
  \begin{minipage}[t]{0.56\linewidth}
\begin{lstlisting}[numbers=none]
qsort([X|L],R) :-
   partition(L,X,L1,L2),
   ( indep(L1, L2) -> 
      qsort(L2,R2) & qsort(L1,R1)
   ; 
      qsort(L2,R2), qsort(L1,R1) ),
   conc(R1,[X|R2],R).
\end{lstlisting}
\figskip
\caption{Parallel QuickSort w/run-time checks.}
\label{fig:qsort-MEL}
  \end{minipage}
\hfill
  \begin{minipage}[t]{0.4\linewidth}
\begin{lstlisting}[numbers=none]


qsort([X|L],R) :-
    partition(L,X,L1,L2),
    qsort(L2,R2) & 
    qsort(L1,R1),
    conc(R1,[X|R2],R).

\end{lstlisting}
\figskip
\caption{Parallel QuickSort.}
\label{fig:qsort-indep-info}
  \end{minipage}
\underfig
\end{figure}

Consider the program in Fig.~\ref{fig:qsort} 
(with \texttt{conc/3} defined as in Fig.~\ref{fig:fnrev}). A possible
parallelization (obtained in this case with the ``MEL''
annotator~\cite{annotators-jlp-short}) is shown in Fig.~\ref{fig:qsort-MEL},
which means that, provided that \kbd{L1} and \kbd{L2} do not have
variables in common at run time, then the recursive calls to
\kbd{qsort} can be run in parallel.
Assuming that \kbd{lt/2} and \kbd{geq/2} in Fig.~\ref{fig:qsort} need
their arguments to be ground (note that this may be either inferred by
analyzing the implementation of \kbd{lt/2} and \kbd{geq/2} or stated
by the user using suitable assertions), the information collected by
the abstract interpreter using, e.g., mode and sharing/freeness
analysis, can determine that \kbd{L1} and \kbd{L2} are ground after
\kbd{partition}, and therefore they do not have variables to share.
As a result, the independence check and the corresponding conditional
is simplified via abstract executability and the annotator yields
instead the code in Fig.~\ref{fig:qsort-indep-info}, which is much
more efficient since it has no run-time check. This check
simplification process is described in detail
in~\cite{effofai-toplas-short} where the impact of abstract
interpretation in the effectiveness of the resulting parallel
expressions is also studied.

The checks in the above example aim at {\em strict} independent
and-parallelism~\cite{sinsi-jlp-short}. However, the annotators are parametrized on the
notion of independence. Different checks can be used for different
independence notions: non-strict independence~\cite{nsicond-sas94-short}, 
constraint-based independence~\cite{consind-toplas-short}, etc.
Moreover, all forms of and-parallelism in logic programs can be seen
as independent and-parallelism, provided the definition of
independence is applied at the appropriate granularity
level.\footnote{For example, stream and-parallelism can be seen as
  independent and-parallelism if the independence of ``bindings''
  rather than goals is considered.}

\ciao~currently includes low-level, native support for the creation of
(POSIX-based) threads at the O.S.\ level which are used as support for
independent and-parallel execution~\cite{hlfullandpar-iclp2008}.  Task
stealing is used to achieve independence between the number of O.S.\
threads and the number of parallel goals~\cite{Hampaper-short,ngc-and-prolog}.

\parbegin
\paragraph{\textbf{Granularity control:}}
the information produced by the \ciaopp\ cost analyzers is also used
to perform combined compile--time/run--time resource control. An
example of this is \emph{task granularity
  control}~\cite{granularity-jsc-short} of parallelized code.  Such
parallel code can be the output of the process mentioned above or code
parallelized manually. In general, this run-time granularity control
process includes computing sizes of terms involved in granularity
control, evaluating cost functions, and comparing the result with a
threshold to decide between parallel and sequential
execution. However, there are optimizations to this general process,
such as cost function simplification and improved term size
computation.

\parbegin
\paragraph{\textbf{Visualization of parallel executions:}}
a tool (VisAndOr~\cite{visandor-iclp93-short}) for depicting parallel
executions was developed and used to help programmers and system
developers understand the program behavior and task scheduling
performed.  This is very useful for tuning the abstract machine and
the automatic parallelizers.

\section{Incremental Compilation and Other Support \\
  for Programming in the Small and in the Large}
\label{sec:incr-comp}

In addition to all the functionality provided by the preprocessor and
assertions, programming in the large is further supported again by the
module system~\cite{ciao-modules-cl2000-short}. This design is the
real enabler of \ciao's modular program development tools, effective
global program analysis, modular static debugging, and module-based
automatic incremental compilation and optimization.  The analyzers and
compiler take advantage of the module system and module dependencies
to reanalyze / recompile only the required parts of the application
modules after one or more of them is changed, automatically and
implicitly, without any need to define ``makefiles'' or similar
dependency-related additional files, or to call explicitly any
``make''-style command.

Application deployment is enhanced beyond the traditional Prolog
top level, since the system offers a full-featured interpreter but
also supports the use of \ciao\ as a scripting language and a compiled
language. Several types of executables can be easily built, from
multiarchitecture \emph{bytecode} executables to single-architecture,
standalone executables.  Multiple platforms are supported, including
the very common Linux, Windows, Mac OS X, and other Un*x-based OSs,
such as Solaris.  Due to the explicit effort in keeping the
requirements of the virtual machine to a minimum, the effort of
porting to new operating systems has so far been reduced.  \ciao~is
known to run on several architectures, including Intel, Power PC,
SPARC, and XScale / ARM processors.

Modular distribution of user and system code in \ciao, coupled with
modular analysis, allows the generation of stripped executables containing
only those builtins and libraries used by the program. Those
reduced-size executables allow programming in the small when strict
space constraints are present.

Flexible development of applications and libraries that use components
written in several languages is also facilitated, by means of compiler
and abstract machine support for multiple bidirectional foreign
interfaces to C/C++, Java, Tcl/Tk, SQL databases (through a notion of
predicate persistence), etc.
The interfaces are described by using assertions, as previously
stated, and any necessary glue code is automatically generated from
them.

\section{An Advanced Integrated Development Environment}
\label{sec:an-advanced-program}

Another design objective of \ciao\ has been to provide a truly
productive program development environment that integrates all of the
tools mentioned before in order to allow the development of correct
and efficient programs in as little time and with as little effort as
possible. This includes a \emph{rich graphical development interface},
based on the latest graphical versions of Emacs and offering menu and
widget-based interfaces with direct access to the top-level/debugger,
preprocessor, and autodocumenter, as well as an embeddable
source-level debugger with breakpoints, and several profiling and
execution visualization tools.  In addition, a plugin with very
similar functionality is also available for the Eclipse programming
environment.\footnote{See
  \texttt{http://eclipse.ime.usp.br/projetos/grad/plugin-prolog/index.html}.}

The programming environment makes it possible to start the top level,
the debugger, or the preprocessor, and to load the current module
within them by pressing a button or via a pair of keystrokes.  
Tracing the execution in the debugger makes the current statement in
the program be highlighted in an additional buffer containing the
debugged file.

\begin{figure}[t]
\centering
\includegraphics[width=0.85\textwidth,trim=0 720 0 0,clip=true]
   {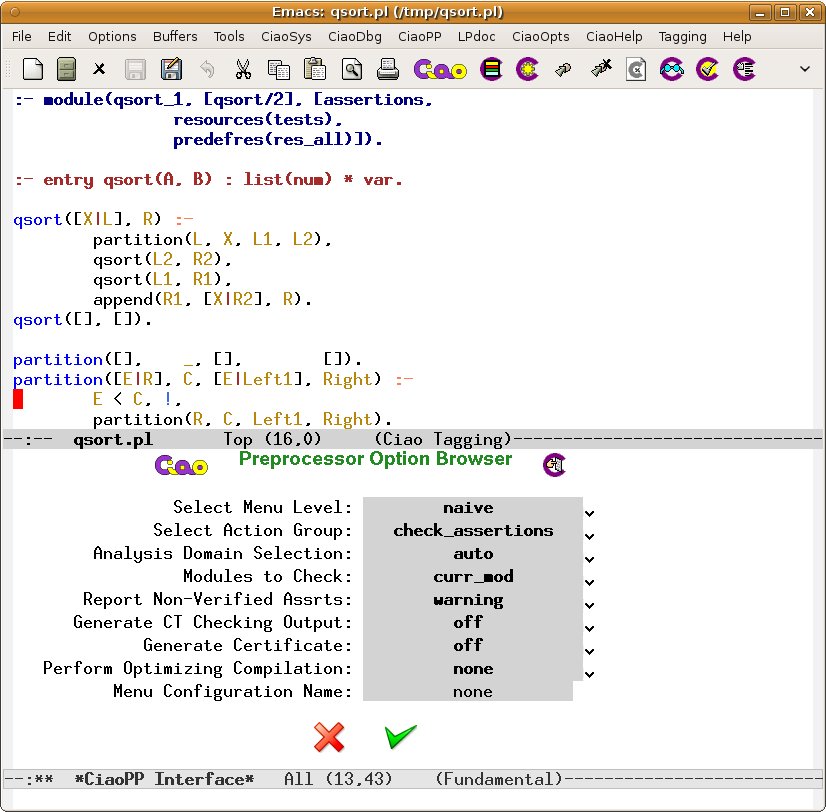}\\ [-0.5mm]
\hspace*{1.15mm}\includegraphics[width=0.85\textwidth,trim=0 0 0 450,clip=true]
   {debugging-ctchecking1a.jpg}
\figskip
  \caption{Menu for compile-time / run-time checking of assertions.}
  \label{fig:debugging-ctchecking1a}
\underfig
\end{figure}

The environment also provides automated access to the documentation,
extensive syntax highlighting, auto-completion, auto-location of
errors in the source, etc., and is highly customizable (to set, for
example, alternative installation directories or the location of some
binaries).
The direct access to the
preprocessor allows interactive control of all the static debugging,
verification, and program transformation facilities.  
For example, 
Fig.~\ref{fig:debugging-ctchecking1a} shows the menu that allows
choosing the different options for compile-time and run-time checking
of assertions 
(this menu is the ``naive'' one, that offers 
reduced and convenient defaults for all others; selecting ``expert''
mode allows changing all options). 

\begin{figure}[t]
  \centering
  \includegraphics[width=\linewidth]{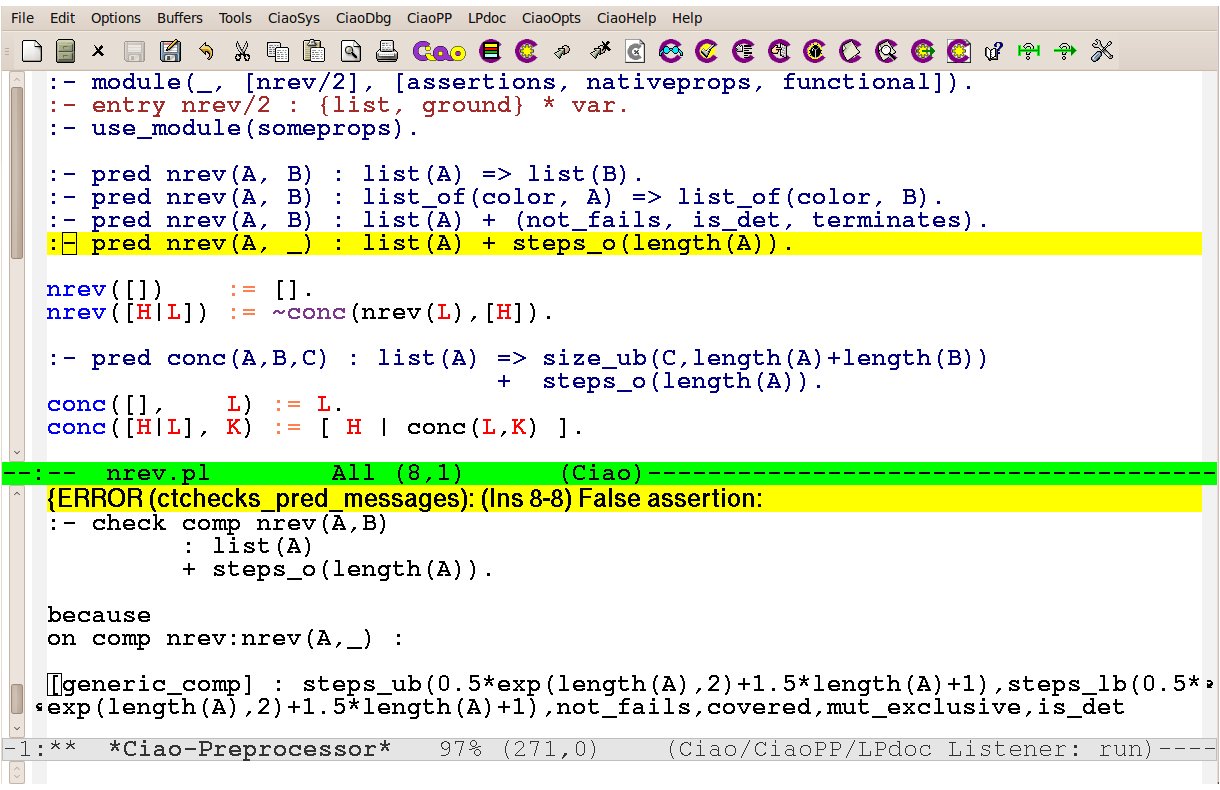}
  \vspace{-2em}
  \caption{Error location in the source --a cost error.}
  \label{fig:semerror}
\underfig
\end{figure}

As another example, Fig.~\ref{fig:semerror} shows \ciaopp\ indicating
a semantic error in the source. In particular, it is the cost-related
error discussed in Section~\ref{sec:static-verif-debug} where the
compiler detects (statically!) that the definition of \texttt{nrev}
does not comply with the assertion requiring it to be of linear
complexity.

The direct access to the auto-documentation
facilities~\cite{lpdoc-cl2000-short} allows the easy generation of
human-readable program documentation from the current file in a
variety of formats from the assertions, directives, and
machine-readable comments present in the program being developed or in
the system's libraries, as well as all other program information
available to the compiler.  This direct access to the documenter and
on a per-module basis is very useful in practice for incrementally
building documentation and making sure that, for example, cross
references between files are well resolved and that the documentation
itself is well structured and formatted.

\section{ Some Final Thoughts: Dynamic vs.\ Static Languages,
  Parallelism}
\label{sec:some-final-thoughts}

We now provide as conclusions some final thoughts regarding how the
now fairly classical \ciao\ approach fares in the light of recent
trends.  We argue that in fact many of the motivations and ideas
behind the development of \ciao\ and \ciaopp\ over the years have
acquired presently even more importance.

The environment in which much software needs to be developed nowadays
(decoupled software development, use of components and services,
increased interoperability constraints, need for dynamic update or
self-reconfiguration, mash-ups) is posing requirements which align
with the classical arguments for dynamic languages but which in fact
go beyond them.  Examples of often required dynamic features include
making it possible to (partially) test and verify applications which
are partially developed, and which will never be ``complete'' or
``final,'' or which evolve over time in an asynchronous, decentralized
fashion (e.g., software service-based systems).
These requirements, coupled with the intrinsic agility in development
of dynamic programming languages such as Python, Ruby, Lua,
JavaScript, Perl, PHP, etc.\ (with Scheme or Prolog also in this
class) have made such languages a very attractive option for a number
of purposes that go well beyond simple scripting. 
Parts written in these languages often become essential components (or
even the whole implementation) of full, mainstream applications.

At the same time, detecting errors at compile time and inferring
properties required to optimize programs are still important issues in
real-world applications.  Thus, strong arguments are also made for
static languages. For example, modern logic and functional languages
(e.g., Mercury~\cite{mercury-jlp-short} or Haskell ~\cite{Haskell2})
impose strong type-related
requirements such as  that all types (and, when relevant, modes) have to
be defined explicitly or that all procedures have to be ``well-typed''
and ``well-moded.''  One argument supporting this approach is that it
clarifies interfaces and meanings and facilitates ``programming in the
large'' by making large programs more maintainable and better
documented.  Also, the compiler can use the static
information to generate more specific code, which can be better in
several ways (e.g., performance-wise).

In the design of \ciao\ we certainly had the latter arguments in mind,
but we also wanted \ciao\ to be useful (as the scripting languages)
for highly dynamic scenarios such as those listed above, for
``programming in the small,'' for prototyping, for developing simple
scripts, or simply for experimenting with the solution to a problem.
We felt that compulsory type and mode declarations, and other related
restrictions, can sometimes get in the way in these contexts.

The solution we came up with involves the rich \ciao\ assertion
language and the \ciao\ methodology for dealing with such
assertions~\cite{aadebug97-informal-short,prog-glob-an-short,assert-lang-disciplbook-short},
which implies making a best effort to infer and check these properties
statically, using powerful and rigorous static analysis tools based on
safe approximations, while accepting that complete verification or
validation may not always be possible and run-time checks may be
needed.  This approach opens up the possibility of dealing in a
uniform way with a wide variety of properties besides types (e.g.,
rich modes, determinacy, non-failure, sharing/aliasing, term
linearity, cost,\ldots), while at the same time making assertions
\emph{optional}.
We argue that this solution has made \ciao\ very useful for
programming both in the small and in the large, combining effectively
the advantages of the strongly typed and untyped language approaches.
In contrast, systems which focus exclusively on automatic compile-time
checking are often rather strict about the properties which the user
can write.  This is understandable because otherwise the underlying
static analyses are of little use for proving the assertions. 

In this sense, the \ciao\ model is related to the \emph{soft typing}
approach~\cite{cartwright91:soft_typing-short}, but without being
restricted to types.  It is also related to the NU--Prolog
debugger~\cite{naish:nu-prolog-debug-89-short}, which performed
compile-time checking of decidable (regular) types and also allowed
calling Prolog predicates at run time as a form of dynamic type
checks.  However, as mentioned before, compile-time inference and
checking in the \ciao~model is not restricted to types (nor requires
properties to be decidable), and it draws many new synergies from its
novel combination of assertion language, properties, certification,
run-time checking, testing, etc.
The practical relevance of the combination of static and dynamic
features is in fact illustrated by the many other languages and
frameworks which have been proposed lately aiming at bringing together
ideas of both worlds.  This includes recent work in gradual typing for
Scheme~\cite{TypedSchemeF08-short} (and the related PLT-Scheme/Racket
language) or Prolog~\cite{schrijvers08:typed_prolog-short}, the recent
uses of ``contracts'' in verification~\cite{clousot}, and the
pragmatic viewpoint of~\cite{lamport99:types_spec_lang}, but applied
to programming languages rather than specification languages.  The
fifth edition of ECMAScript, on which the JavaScript and ActionScript
languages are based, includes optional (soft-)type declarations to
allow the compiler to generate more efficient code and detect more
errors.  The Tamarin project~\cite{mozilla:tamarin} intends to use
this additional information to generate faster code.  The
RPython~\cite{Ancona-07-short} language imposes constraints on the
programs to ensure that they can be statically typed.  RPython is
moving forward as a general purpose language.  This line has also
brought the development of safe versions of traditional languages,
such as, e.g., CCured \cite{DBLP:journals/toplas/NeculaCHMW05-short}
or Cyclone~\cite{DBLP:conf/usenix/JimMGHCW02} for C, as well as of
systems that offer capabilities similar to those of the \ciao\
assertion preprocessor, such as Deputy
(\url{http://deputy.cs.berkeley.edu/}) or
Spec\#~\cite{DBLP:journals/fac/LeavensLM07}.

We believe that \ciao\ has pushed and is continuing to push the state
of the art in solving this currently very relevant and challenging
conundrum between statically and dynamically checked languages. It
pioneered what we believe is the most promising approach in order to
be able to obtain the best of both worlds: the combination of a
flexible, multi-purpose assertion language with strong program
analysis technology.  This allows support for dynamic language
features while at the same time having the capability of achieving the
performance and efficiency of static systems. 
We believe that a good part of the power of the \ciao\ approach also comes
from the synergy that arises from using the same framework and
assertion language for different tasks (static verification, run-time
checking, unit testing, documentation, \ldots) and its interaction
with the design of \ciao\ itself (its module system, its
extensibility, or the support for predicates and
constraints).
The fact that properties are written in the source language is
instrumental in allowing assertions which cannot be statically
verified to be translated easily into run-time checks, and this is
instrumental in turn in allowing users to get some benefits even if a
certain property cannot be verified at compile time.
The assertion language design also allows a smooth integration with
unit testing.  Moreover, as (parts of) the unit tests that can be
verified at compile time are eliminated, sometimes unit tests can be
checked without ever running them.

Another interesting current trend where \ciao's early design choices
have become quite relevant is parallelism.  Multi-core processors are
already the norm, and the number of cores is expected to grow in the
foreseeable future. This has renewed the interest in language-related
designs and tools which can simplify the intrinsically
difficult~\cite{Karp88} but currently necessary task of parallelizing
programs.  In the \ciao\ approach programmers can choose between
expressing manually the parallelism with high-level constructs,
letting the compiler discover the parallelism, or a combination of
both.  The parallelizer also checks manual parallelizations for
correctness and, conversely, programmers can easily inspect and
improve the (source level) parallelizations produced by the
compiler. These capabilities rely (again) on the use of \ciaopp's
powerful, modular, and incremental abstract interpretation-based
static program analyzers. This approach was pioneered by \&-Prolog and
\ciao\ (arguably one of the first direct uses of abstract
interpretation in a real compiler), and seems the most promising
nowadays, being adopted by many systems (see,
e.g.,~\cite{tutorial-europar97-short} for further discussion).

\parbegin
\vspace*{10mm}
\paragraph{\textbf{Probing Further.}}
\label{sec:probing-further}

The reader is encouraged to explore the system, its documentation, and
the tutorial papers that have been published on it. At the time of
writing, work is progressing on the new 1.14 system version which
includes significant enhancements with respect to the previous major
release (1.10).
In addition to the autodocumenter, new versions also include within
the default distribution the \ciaopp\ preprocessor (initially beta
versions), which was previously distributed on demand and installed
separately.  The latest version of \ciao, 1.13, which is essentially a
series of release candidates for 1.14 has now been available for some
time from the \ciao\ web site (snapshots) and subversion repository.

\parbegin
\paragraph{\textbf{Contact / download info / license:}} the latest
versions of \ciao~can be downloaded from \url{http://www.ciaohome.org}
or \url{http://www.cliplab.org}. \ciao~is free software protected to
remain so by the GNU LGPL license, and can be used freely to develop
both free and commercial applications.

\begin{small}
\parbegin
\paragraph{\textbf{Acknowledgments:}} 
The \ciao\ system is in continuous and very active development through
the collaborative effort of numerous members of several institutions,
including UPM, the IMDEA Software Institute, UNM, UCM, Roskilde U.,
U.\ of Melbourne, Monash U., U.\ of Arizona, Link\"{o}ping U., NMSU,
K.\ U.\ Leuven, Bristol U., Ben-Gurion U., INRIA, as well as many
others.  The development of the \ciao\ system has been supported by a
number of European, Spanish, and other international projects;
currently by the European IST-215483 \emph{S-CUBE} and FET IST-231620
{\em HATS} projects, the Spanish 2008-05624/TIN \emph{DOVES} project,
and the CAM P2009/TIC/1465 \emph{PROMETIDOS} project.
Manuel Hermenegildo was also supported previously by the
IST Prince of Asturias Chair at the University of New Mexico.  The
system documentation and related publications contain more specific
credits to the many contributors to the system.
We would also like to thank the anonymous reviewers and the editors of
the special issue for providing very constructive and useful comments
which have greatly contributed to improving the final version of the paper.
\end{small}

\vspace*{-3mm}
\begin{small}
\bibliographystyle{acmtrans}

\end{small}

\end{document}